\documentclass[sn-mathphys-num]{sn-jnl}

\usepackage{graphicx}%
\usepackage{multirow}%
\usepackage{amsmath,amssymb,amsfonts}%
\usepackage{amsthm}%
\usepackage{mathrsfs}%
\usepackage[title]{appendix}%
\usepackage{xcolor}%
\usepackage{textcomp}%
\usepackage{manyfoot}%
\usepackage{booktabs}%
\usepackage{algorithm}%
\usepackage{algorithmicx}%
\usepackage{algpseudocode}%
\usepackage{listings}%
\usepackage{tabularx}
\usepackage{mathtools}
\usepackage[thinc]{esdiff}
\usepackage{float}

\theoremstyle{thmstyleone}%
\theoremstyle{thmstyletwo}%

\theoremstyle{thmstylethree}%

\begin{document}

\title[Article Title]{Symmetry and Critical Dynamics in Supercooled Liquid Crystals: Insights into the Glass Transition}


\author*[1,2]{\fnm{Szymon} \sur{Starzonek}}\email{szymon.starzonek@uj.edu.pl}

\author[3]{\fnm{Ale\v{s}} \sur{Igli\v{c}}}\email{ales.iglic@fe.uni-lj.si}

\author[4]{\fnm{Aleksandra} \sur{Drozd-Rzoska}}\email{arzoska@unipress.waw.pl}

\author[4]{\fnm{Sylwester J.} \sur{Rzoska}}\email{sylwester.rzoska@unipress.waw.pl}

\affil*[1]{\orgdiv{Institute of Theoretical Physics}, \orgname{Jagiellonian University}, \orgaddress{\street{\L ojasiewicza 11}, \city{Krak\'{o}w}, \postcode{30-348}, \country{Poland}}}

\affil[2]{\orgdiv{Mark Kac Center for Complex Systems Research}, \orgname{Jagiellonian University}, \orgaddress{\street{\L ojasiewicza 11}, \city{Krak\'{o}w}, \postcode{30-348}, \country{Poland}}}

\affil[3]{\orgdiv{Laboratory of Physics, Faculty of Electrical Engineering}, \orgname{University of Ljubljana}, \orgaddress{\street{Tr\v{z}a\v{s}ka 25}, \city{Ljubjana}, \postcode{1000}, \country{Slovenia}}}

\affil[4]{\orgname{Institute of High Pressure Physics of the Polish Academy of Sciences}, \orgaddress{\street{Soko\l owska 29/37}, \city{Warszawa}, \postcode{01-147}, \country{Poland}}}


\abstract{This study introduces a modeling approach aimed at elucidating the pivotal role of symmetry in phase transitions, focusing specifically on the isotropic-nematic (I-N) transition characteristic of liquid crystal systems. By leveraging insights from the Ising model and incorporating considerations of topological defects, the transition to the glassy state in rod-like molecular systems in the supercooled state is examined. Through a critical-like analysis of the system's dynamical properties, universality classes directly linked to symmetry are discerned. This paper delves into the role of symmetry in the glass transition, as manifested in the generalized critical relation of configurational entropy $S_C(T)=S_0(1?T_K/T)^n$, where the critical exponent $n$ is intricately tied to the system's symmetry. The determined values of the pseudocritical exponent $n$ exhibit universality across the studied systems and demonstrate excellent agreement with thermodynamic data. Furthermore, the congruence between the dynamic representation, as indicated by the primary relaxation time, and the thermodynamic representation, exemplified by the specific heat capacity, underscores the robustness of the findings. The identification of critical-like behavior and the observation of symmetry breaking during the transition to the glass state suggest its intrinsic thermodynamic nature. This work provides a unified framework for understanding the glass transition, bridging dynamic and thermodynamic perspectives through the lens of symmetry.}




\maketitle

\section{Introduction}\label{sec1}

Soft matter systems display a rich array of phase transitions, often defying simple macroscopic descriptions due to weakly discontinuous transitions or the occurrence of numerous correlated transformations along similar temperature and pressure profiles. In such cases, the inherent symmetry of the physical systems emerges as a crucial determinant, playing a pivotal role in their behavior. Leveraging the methodology of statistical mechanics, particularly in the context of critical phenomena, enables the determination of order parameters in soft matter phases, thereby facilitating the characterization of symmetry in phase transition processes.  A fundamental inquiry lies in identifying physical quantities, both dynamic and static, exhibiting critical-like behavior near phase transition points, with particular emphasis on their correlation with system symmetry or symmetry changes. The breaking of symmetry during phase transitions, particularly in soft matter systems, plays a crucial role in determining the system's behavior and the nature of the glass transition \cite{bib1, bib2, bib3}

The transition to the glassy state remains a compelling challenge for 21st-century physics, characterized by the intricate interplay between order and disorder, better understood through the lens of symmetry and its breaking. For instance, ferroic systems exhibit local symmetry breaking at the glass transition temperature $T_g$. Early attempts by Angel to correlate symmetry breaking with macroscopic system descriptors, such as configurational entropy, laid important groundwork, introducing the concept of the configuron to delineate distinct system states \cite{bib4}.

The notion of configurational entropy $S_C(T)$ assumes critical significance in glass physics, yet its treatment remains nebulous and imprecise. Traditional approaches, such as deriving the Vogel-Fulcher-Tammann (VFT) equation via substitutions into the Adam-Gibbs relation and assuming $T_K=T_0$, have failed to offer a coherent model, resulting instead in a multi-parameter function that sidelines the role of configurational entropy. Despite extensive research efforts spanning simulations, theoretical calculations, and experimental observations, a unified model for the glass transition remains elusive. The convergence between dynamic and thermodynamic characteristics observed experimentally underscores the need for theoretical frameworks linking the glass transition temperature $T_g$ with latent phase transitions. Experimental data and theoretical predictions regarding the thermodynamic and dynamic nature of the ultraviscous region, along with the widespread use of the VFT equation, underscore the relevance of configurational entropy in determining $T_K$, as expressed in Equation (1).
\begin{equation}
	S_C (T)=S_0 \left(1-\frac{T_K}{T}\right)=S_0 t
\end{equation}
where $S_0$ is an entropy of a liquid at high temperature, $T_K$ is a  Kazumann temperature and $t=(1-T_K/T)$.
In our last paper \cite{bib5} we introduced a critical-like relation for configurational entropy. It based on calculations of specific heat capacities between supercooled liquid and glassy state. Parametrised relation may be expressed by relation
\begin{equation}
	S_C (T)=S_0 t^n
\end{equation}
where $n$ is an introduced pseudo-critical exponent.
A notable characteristic of configurational entropy $S_C(T)$ is its marked decline in proximity to the glass transition, a phenomenon that engenders a notable paradox within the field \cite{bib3,bib5,bib6,bib7,bib8,bib9,bib10,bib11,bib12,bib13,bib14,bib15,bib16,bib17,bib18,bib19}. This enigma arises from the presumption that the reduction in temperature ($T_g$) should lead to a corresponding decrease in entropy, potentially resulting in a scenario where the extrapolated entropy of the liquid phase falls below that of the crystalline phase ($S_{liq}<S_{cr}$), implying a negative configurational entropy ($S_C(T)<0$). Such a scenario is manifestly untenable in physical terms, leading to the identification of a critical temperature $T(S_C=0)$, often referred to as the Kauzmann temperature $T_K$ \cite{bib2,bib3,bib4}. The entropy crisis, commonly known as the Kauzmann paradox, revolves precisely around this issue of negative configurational entropy below $T_g$. The Kauzmann paradox arises when the extrapolated entropy of a supercooled liquid falls below that of the crystalline phase, suggesting a negative configurational entropy, which is physically untenable. This paradox highlights the need for a deeper understanding of the glass transition. It is claimed that the formation of glass serves as a resolution to this quandary by infusing entropy into the system precisely at the juncture of the glass transition, thereby entrapping it in an amorphous state \cite{bib3,bib4,bib5,bib6,bib7,bib8,bib9,bib10,bib11,bib12,bib17,bib18,bib19}. This crisis yields crucial insights into the determination of $T_g$, suggesting that under conditions of arbitrarily slow cooling, $T_K$ represents the ultimate limiting value towards which $T_g$ converges ($T_g \rightarrow T_K$ as $\dot{T}=0$).

In 1965, Adam and Gibbs (AG), postulating the cooperative nature of molecular reorientation, introduced the concept of cooperatively rearranging regions (CRR), defining them as the smallest volumes capable of independent configuration changes amidst adjacent areas \cite{bib3,bib4,bib5,bib6,bib7,bib8,bib9,bib10,bib11,bib12}. Leveraging insights from the theory of critical phenomena \cite{bib14,bib15,bib16}, they inferred that as temperature decreases, the correlation radius of CRRs expands. Given that the activation energy is proportionate to the volume of CRRs ($E_a\propto \xi_{CRR}^3$), Adam and Gibbs delineated the relationship between relaxation time and the number of molecules enclosed within cooperatively rearranged regions:
\begin{equation}
	\tau(T)=\tau_{AG}\exp{\left(\frac{A_{AG}}{TS_C(T)} \right)}
\end{equation}
where $A_{AG}$ is the energy barrier between the states, $\tau_{AG}$ is the coefficient related to the high temperature dynamic region, and $S_C (T)$ is the configuration entropy assuming that $N(T)\approx S_C (T)$. 

From a thermodynamic perspective, the configurational entropy can be linked to the variation in specific heat at the onset of the glass transition, denoted as $\Delta C_p^g$. Assuming a model where $\Delta C_p^g\approx A/T$ yields the Vogel-Fulcher-Tammann (VFT) relation for the temperature-dependent relaxation time. This formulation enables the estimation of the size of cooperatively ordered regions through the expression $N(T) \approx [A(T-T_0)]^{-1}$, where $T_0$ signifies the temperature characterizing the theoretical glass transition.

Utilizing such systems facilitates the modeling of behaviors as they approach the glass transition temperature. The rod-like structure of molecules, coupled with the absence of significant intermolecular interactions, allows for a focus on critical-like dynamics near $T_g$. Conversely, by examining various supercooled phases (such as isotropic, nematic, and chiral nematic), universality in phase behaviors can be discerned. The robust experimental validation of these systems, alongside comparisons with substances like glycerol, which feature nearly spherical molecules, adds further value to the derived outcomes \cite{bib5,bib17,bib18,bib19}.

Although the Adam-Gibbs model does not directly address the absolute size of cooperatively rearranging regions (CRRs) at the glass transition temperature $T_g$, its utility has been widely acknowledged across numerous experimental domains owing to its ability to seamlessly integrate dynamic and thermodynamic properties. Furthermore, it offers insights into the pronounced increase in relaxation time as temperature decreases, contributing to the depiction of the glass transition as a largely continuous process. Thus, the Adam-Gibbs model serves as a robust foundation for adopting a critical-like approach to describing glass transition phenomena.

\section{Materials and methods}\label{sec2}

The liquid crystals (8*OCB, E7 mixture, 5*CB) were synthesized and purified at the Military University of Technology (Warsaw, Poland), ensuring high purity for experimental measurements.

The liquid crystals 8*OCB and 5*CB exhibit a chiral nematic phase characterized by their rod-like molecular structure. Similarly, E7 is a liquid crystal mixture comprising compounds such as 5CB, 7CB, 8OCB, and 5CT, all of which share a rod-like molecular structure.

The heat capacity data, crucial for configurational entropy calculations, were extracted from Ref. \cite{bib5}.

Dielectric measurements were conducted using an alpha-A impedance analyzer (Novocontrol, Germany) across a frequency range of $10^{9}$ Hz to $10^{-2}$ Hz and a temperature range of 323 K to 123 K. Samples were placed in flat, round parallel capacitors made of steel with a diameter of 20 mm and a gap of 0.2 mm, separated by a Teflon ring. Temperature stabilization was achieved using a Quatro CryoSystem (Novocontrol, Germany) with liquid nitrogen (N2), ensuring a stability better than $\delta T = 0.02$ K.

The supercooled state was attained by gradually cooling the samples following annealing at 323 K with a cooling rate of $\dot{T}=10$ K/min. The studied liquid crystals manifest in supercooled isotropic (Iso), nematic (N), and chiral nematic (N*) phases for 8*OCB, E7 mixture, and 5*CB, respectively. As a comparative reference, supercooled glycerol was also investigated.

\section{Results and discussion}\label{sec3}
Figure 1 shows the configurational entropy $S_{C}(T)$ for all tested systems, calculated from heat capacity data. The dashed lines represent fits using Eq. (1) with \(n = 1\), while the solid lines represent fits using Eq. (2) with \(n \neq 1\). Table 1 compares the values of the exponent \(n\) obtained from configurational entropy and relaxation time data, showing excellent agreement between the two methods. The data obtained from the analysis of the configurational entropy have been collected in Table 1. Their values agree with the values of the parameter n obtained earlier in the dynamic tests of the relaxation time with the accuracy of the measurement error \cite{bib1,bib13} and Ref. therein]. The obtained results suggest that the tested liquid crystals show $n>1$, while for glycerol $n=1$. It is postulated that the parameter $n$ is related to the symmetry of the system \cite{bib17,bib18,bib19}. Many experimental data confirm this hypothesis \cite{bib17,bib18}. In the case of liquid crystal materials, there is a uniaxial symmetry assigned to the value n>1. On the other hand, spherical symmetry corresponds to $n=1$, while $n<1$ is encountered in the plastic phase, in which positional symmetry occurs \cite{bib17} and Ref. therein. The pseudocritical exponent \(n\) is closely related to the symmetry of the system, with \(n > 1\) observed in liquid crystals due to their uniaxial symmetry, while \(n = 1\) corresponds to spherical symmetry in systems like glycerol.

\begin{figure}[h!]
	\centering
	\includegraphics[width=0.8\textwidth]{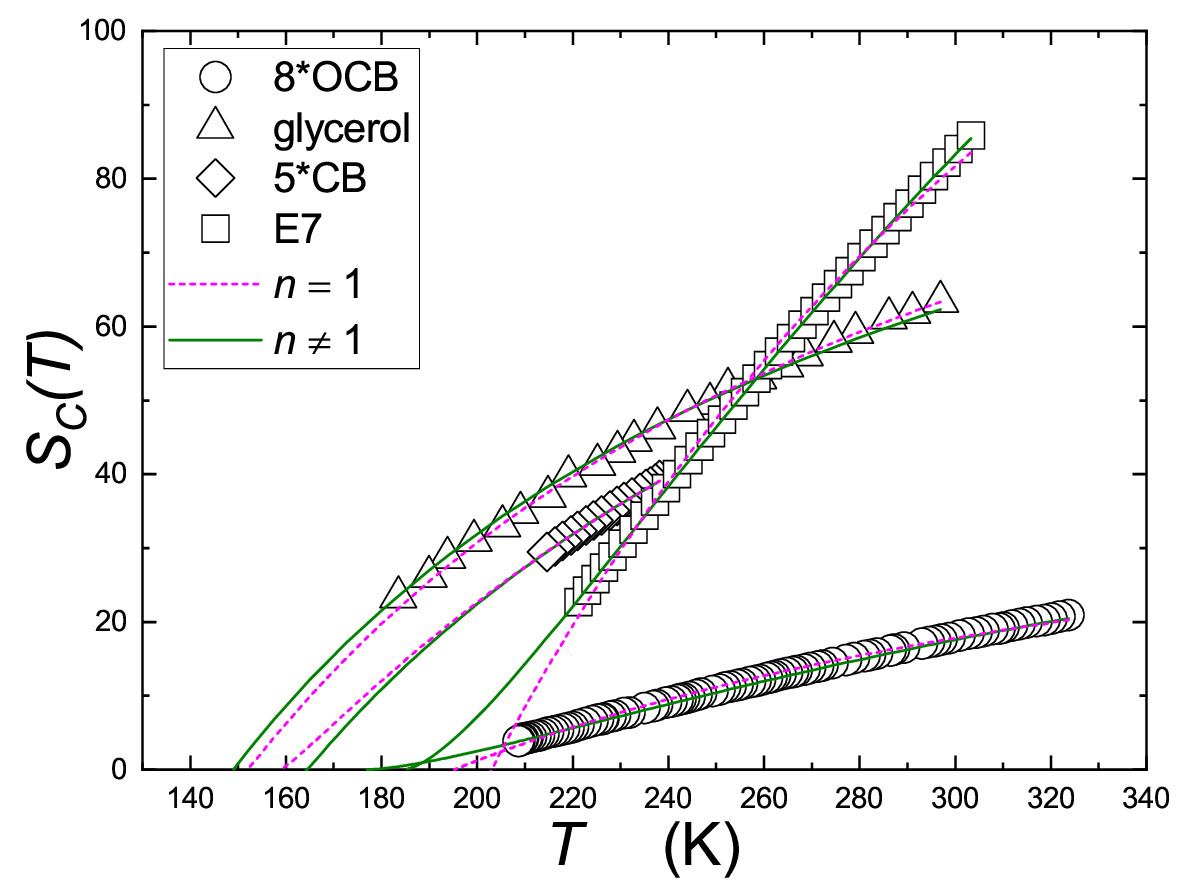}
	\caption{Configurational entropy $S_C(T)$ for studied systems calculated from heat capacities data (see Ref. [5]). The Dashed lines denote fitting by Eq. (1) with $n=1$, whereas solin ones by Eq.(2) with $n\neq 1$. }\label{fig1}
\end{figure}
\begin{table}[h]
	\caption{Comparison of exponents $n$ calculated on the basis of relaxation times and configurational entropy}\label{tab:exponents}%
	\begin{tabularx}{0.8\textwidth}{@{}>{\centering\arraybackslash}X >{\centering\arraybackslash}X >{\centering\arraybackslash}X@{}}
		\toprule
		\textbf{System} & \textbf{$n$ \newline from $S_C(T)$}  & \textbf{$n$ \newline from $\tau(T)$} [18] \\
		\midrule
		E7  & 1.51  & 1.53  \\
		5*CB  & 1.53 (iso)  & 1.49 (iso) \\
		8*OCB  & 1.49  & 1.53  \\
		glycerol  & 1.02  & 1.00  \\
		\bottomrule
	\end{tabularx}
\end{table}
The basic method of studying the distribution of relaxation times is the differential analysis of the basic equations describing the dynamics. Due to the linearisation of data, it becomes possible to identify dynamic domains and determine the parameters of the data-fitting function.
\begin{figure}[h!]
	\centering
	\includegraphics[width=0.8\textwidth]{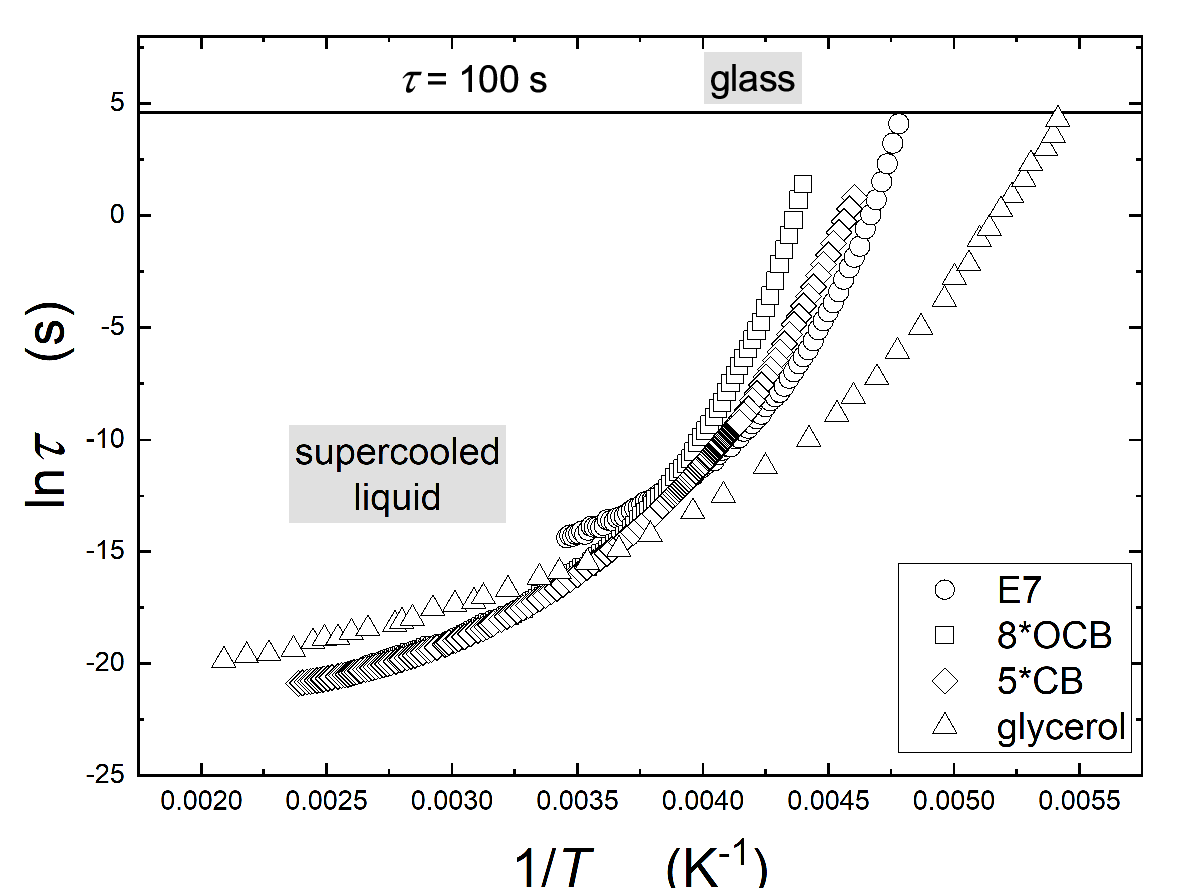}
	\caption{Temperature dependence of the primary relaxation time $\tau(T)$ for studied systems. The glass transition temperature was obtained for $\tau(T_g)\approx100$ s.}\label{fig2}
\end{figure}

Fig. 2 shows the dependence of the relaxation time distribution on the temperature for all tested systems. Due to the primary relaxation process in the ultraviscous (supercooled) region, the typical Arrhenius behaviour does not occur.
According to the dynamic-thermodynamic description of the supercooled region, it is possible to substitute the generalized configuration entropy equation (2) to the Adam-Gibbs relation (3), which allows for the scaling procedure, as follows
\begin{equation}
	\tau(T)=\tau_{AG}\exp{\left(\frac{A_{AG}}{TS_0t^n} \right)}
\end{equation}
and then
\begin{equation}
	\ln \tau(T)=\ln? \tau_{AG} + \frac{A_{AG}}{TS_0\left(\frac{T-T_K}{T}\right)^n }
\end{equation}

However, one assumption should be made, namely that the Kauzmann temperature value was determined from the fit of the configuration entropy. The above procedure will allow to check whether during the analysis of dynamic data (primary relaxation time) it is possible to obtain the value of the exponent n identical for the measurements of the specific heat capacity $C_p (T)$.

As a result a following scaling procedure was performed. Taking and derived Eq.(5) one gets
\begin{equation}
	\diff{\ln \tau(T)}{T}=-\frac{A_{AG}((n-1)T_K+T)\left( \frac{T-T_K}{T}\right)^{-n}}{S_0T^2(T-T_K)}
\end{equation}
and substituting $A^{'}=A_{AG}/S_0$ it gives
\begin{equation}
	-T^3\diff{\ln \tau(T)}{T}=A^{'}((n-1)T_K+T)\left( \frac{T-T_K}{T}\right)^{-n-1}
\end{equation}
after assuming $B^{'}=(n-1)T_K$, taking logarithm and simplifying we obtain linearised relation, as follows
\begin{equation}
	\ln \left[-T^2 \diff{\ln \tau(T)}{T} \right]=\ln A^{'}  +(-n-1) \ln \left[ \frac{T-T_K}{T}\right] = ax+b
\end{equation}
where for $n=1$ the slope equals $a=-2$.
\begin{figure}[h!]
	\centering
	\includegraphics[width=1\textwidth]{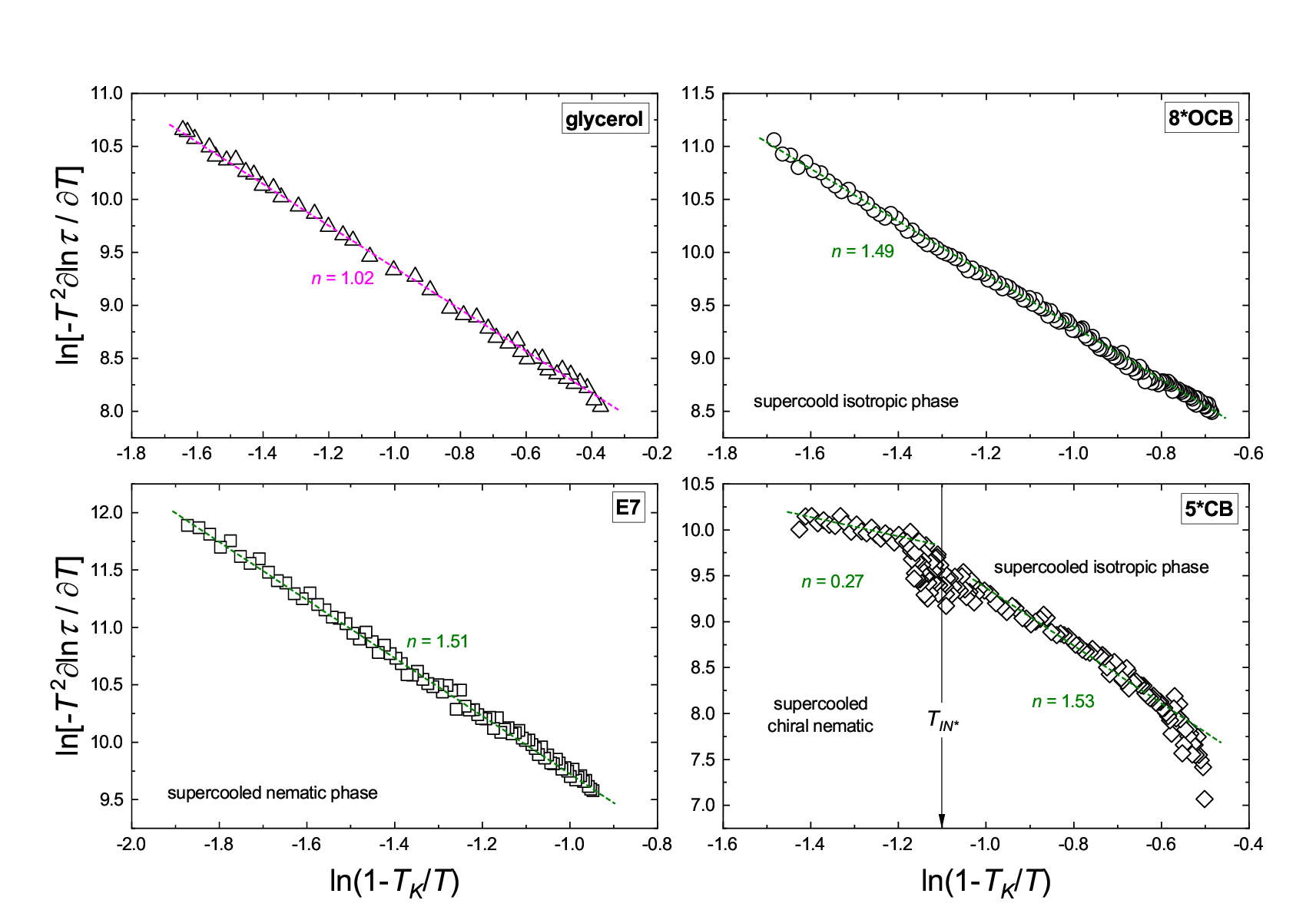}
	\caption{Scaling of the Adam-Gibbs relation by the use of critical-like description based on configurational entropy definition given by Eq.(8). The slope denotes the pseudo-critical exponent $n=-a-1$. The Kauzmann temperatures obtained from Fig.1 and Eq.(1) are also presented.}\label{fig3}
\end{figure}

Fig. 3 shows the performed scaling of the primary relaxation time described by the Adam-Gibbs equation (4) for the tested liquid crystal systems and glycerol. Because of linearisation, it was possible to determine the exponent $n = b-1$ from the slope of data. The obtained results are consistent with the values obtained on the basis of specific heat data with the accuracy of the error resulting from the numerical calculation methods \cite{bib5}. It is worth mentioning that this is the first attempt of this type to combine the dynamic description (relaxation time) and the thermodynamic description (configuration entropy) in glass-forming systems.

The key to the above research is to obtain satisfactory experimental data near the glass transition with very high accuracy. This applies to both specific heat and relaxation processes. The choice of systems to be analysed is not without significance. Because of the rod-like shape and the lack of intermolecular interactions, a model approach to the problem of the glass transition is possible. It is also worth emphasising the Debye-like shape of the absorption curve near $T_g$, what in relation to molecular structure may provide to universal behaviour in ultraviscous (supercooled) dynamic region \cite{bib5,bib17,bib18,bib19}.

\section{Conclusions}\label{sec4}
Our study demonstrates that the configurational entropy \(S_{C}(T)\), calculated from specific heat capacity data, can be described using a critical-like relation with the pseudocritical exponent \(n\). This exponent, which reflects the system's symmetry, shows excellent agreement between dynamic (relaxation time) and thermodynamic (specific heat capacity) data \cite{bib5,bib17,bib18,bib19}. The observed critical-like behavior and symmetry breaking during the glass transition suggest its intrinsic thermodynamic nature. These findings provide a unified framework for understanding the glass transition, bridging dynamic and thermodynamic perspectives and offering new insights into the role of symmetry in phase transitions.

The critical-like nature of configurational entropy gives it a completely new meaning in the physics of the amorphous phase and the description of previtrification dynamics. Its confirmation may be the characteristic behaviour of the specific heat just before $T_g$ \cite{bib5}, which leads to the question of the existence of thermodynamic pretransitional effects related to critical fluctuations.

The scaling of the Adam-Gibbs relation allows to determine the dynamic parameter $n$, related to the symmetry of the system, from the slope of the straight line, a value equals to that obtained from the thermodynamic data. The above conclusions suggest the thermodynamic nature of the transition to the glass state.

\section*{Data availability statement}
The authors declare that the data supporting the findings of this study are available within the paper and its supplementary information files.

\bibliography{sn-article}
\end{document}